\documentclass[aps,10pt,reprint,superscriptaddress,showpacs]{revtex4-2}
\usepackage{xcolor}
\usepackage{physics}
\usepackage{amssymb}
\usepackage{graphicx}
\usepackage{mathtools}
\usepackage{amsmath}
\usepackage{hyperref}
\usepackage{dsfont}

\usepackage{amsfonts}

\newcommand{\KITP}{\affiliation{Kavli Institute for Theoretical Physics, University of California, Santa Barbara, CA 93106, USA}}
\newcommand{\TUM}{\affiliation{Technical University of Munich, TUM School of Natural Sciences, Physics Department, 85748 Garching, Germany}}
\newcommand{\MCQST}{\affiliation{Munich Center for Quantum Science and Technology (MCQST), Schellingstr. 4, 80799 M{\"u}nchen, Germany}}

\begin{document}
\title{Detecting and stabilizing measurement-induced symmetry-protected topological phases in generalized cluster models}

\author{Raúl Morral-Yepes} \email{raul.morral@tum.de} \TUM \MCQST 
\author{Frank Pollmann} \TUM \MCQST
\author{Izabella Lovas} \TUM \KITP

\begin{abstract}
We study measurement-induced symmetry-protected topological (SPT) order in a wide class of quantum random circuit models by combining calculations within the stabilizer formalism with tensor network simulations. We construct a family of quantum random circuits, generating the out-of-equilibrium version of all generalized cluster models, and derive a set of nonlocal string order parameters to distinguish different SPT phases. We apply this framework to investigate a random circuit realization of the XZX cluster model, and use the string order parameter to demonstrate that the phase diagram is stable against extending the class of unitary gates in the circuit, from Clifford gates to Haar unitaries. We then turn to the XZZX generalized cluster model, and demonstrate the coexistence of SPT order and spontaneous symmetry breaking, by relying on string order parameters and a connected correlation function.
\end{abstract}

\maketitle

\section{Introduction}

Topology in quantum many-body systems has been at the forefront of condensed matter research in recent years~\cite{quantum_info_meets_quantum_matter}. Topological invariants allow us to classify the ground states of gapped local Hamiltonians into distinct phases~\cite{ent_spectrum_Pollmann, RevModPhys.88.035005,RevModPhys.89.041004,PhysRevB.83.035107, PhysRevB.84.235128}, characterized by nonlocal order parameters, and displaying exotic properties, such as anyonic excitations or gapless edge states. A particularly rich phase diagram arises in the presence of symmetries, displaying various symmetry-protected topological (SPT) phases that cannot be characterized in terms of the spontaneous breaking of a global symmetry~\cite{science.1227224, annurev-conmatphys-031214-014740}. Instead, they are characterized by entanglement patterns between subsystems~\cite{ent_spectrum, ent_spectrum_Pollmann}, captured by a topological entanglement entropy~\cite{topent_PhysRevLett.96.110404, topo_ent_entropy}, as well as nonlocal ``string order''~\cite{string_order, detection_spt_string_order}.

Recently, the concept of SPT phases has been extended from equilibrium systems to nonequilibrium scenarios~\cite{Lavasani2021} in the context of measurement-induced entanglement transitions in quantum random circuits~\cite{review_random_circuits, EPT1,Li_2019,  EPT2, EPT3}. In these quantum circuits, the time evolution is governed by a competition between random unitary gates, spreading information and tending to scramble the system, and repeated local measurements, reducing entanglement. The interplay of these opposing effects leads to dynamical phase transitions between different stationary states:  a highly entangled thermal state characterized by a volume law scaling of subsystem entanglement entropy, and nonthermal area law states~\cite{review_random_circuits, EPT1,Li_2019,  EPT2, EPT3, A5, A7, EPT4, A3, A1, A6, A8, A9, A10, A11, A13, MIPT1, MIPT2, A2, A12, MIPT3,Khemani_PhysRevLett.126.060501,Grover_PRXQuantum.2.040319, Sierant_2022, zerba2023measurement, PhysRevB.106.024305, MIPT4, MIPT5, MIPT6, Khemani_PhysRevX.12.011045}, with different types of measurements generating novel nonequilibrium phases of matter~\cite{Lavasani2021, Sang_2021, Klocke_2022, Lavasani2, Lavasani3, Wampler_2022, SPT_measurement_only, Kells, Bardarson}. In particular, in Ref.~\cite{Lavasani2021}, the authors have demonstrated that the area law stationary state can also exhibit SPT order, similarly to the area entangled ground states of gapped Hamiltonians.

These recent advances raise exciting questions about the measurement-induced topological phases in quantum circuits. The SPT phase found in Ref.~\cite{Lavasani2021} emerged in a Clifford quantum random circuit model in the presence of a protecting $\mathbb{Z}_2\times\mathbb{Z}_2$ symmetry, and was detected through a topological entanglement entropy. Generalizing this construction to other types of topological phases, as well as finding and classifying the topological phases accessible in out-of-equilibrium systems, can offer new insights into the properties of dynamical phase transitions. Another interesting aspect concerns the order parameter of the phase transition. Clifford random circuits have a special structure, allowing us to simulate them efficiently by relying on the stabilizer formalism~\cite{AaronsonGottesman}. In this setting, the topological entanglement entropy is well suited for detecting SPT order in the numerical calculations. However, the topological entanglement entropy is very difficult to access in experimental realizations. For this reason, it is important to identify other, more accessible order parameters. Another open question is the stability of the SPT phase in the wider class of Haar random circuits.

In this paper, we take the first steps towards answering these questions. We generalize the construction of Ref.~\cite{Lavasani2021} to generate the whole family of generalized cluster models~\cite{generalized_cluster_models, Verresen17}. This extended set of random circuits hosts different types of SPT phases, as well as phases with simultaneous SPT order and spontaneous symmetry breaking (SSB). We also construct a set of nonlocal string order parameters and demonstrate that they are capable of distinguishing the different phases realized by the circuits, thereby providing a convenient alternative to topological entanglement entropy that is more accessible both numerically and experimentally~\cite{Sompet_2022}. To this end, we analyze two members of the family of generalized cluster models in detail, by combining simulations in the stabilizer formalism with tensor network methods. First, we focus on the XZX model already examined in Ref.~\cite{Lavasani2021}, and confirm that the string order parameter can be used to determine the full phase diagram. By relying on tensor network simulations, we also show that the phase diagram is remarkably stable against extending the class of unitaries from random Clifford to random Haar, provided that the protecting $\mathbb{Z}_2\times\mathbb{Z}_2$ symmetry is respected by the gates. Secondly, we consider the so-called XZZX cluster model and demonstrate that it hosts a phase with coexisting SPT order and SSB. We determine the full phase diagram of this model within the stabilizer formalism, by evaluating the string order parameters and a connected correlator capturing SSB.

The paper is organized as follows. We first discuss the general theoretical framework in Sec.~\ref{sec:framework}. Here, we introduce a set of quantum random circuits, realizing an out-of-equilibrium version of the whole family of generalized cluster models. We also construct string order parameters capturing the SPT order in the area law stationary states of these circuits. We then turn to the XZX cluster model in Sec.~\ref{sec:xzx}. First, we focus on Clifford random circuits in Sec.~\ref{subsec:clifford}, and we validate the string order parameter proposed before, by using it to obtain the full phase diagram and comparing it to predictions relying on entanglement entropies from Ref.~\cite{Lavasani2021}. We test the stability of this phase diagram by extending the class of random unitary gates from Clifford to Haar random gates in Sec.~\ref{subsec:haar}. We then demonstrate the coexistence of SPT order with spontaneous symmetry breaking by studying the XZZX cluster model in Sec.~\ref{sec:xzzx}. We summarize our main conclusions in Sec.~\ref{sec:conclusion}.

\section{General framework}\label{sec:framework}

In this section, we construct a family of quantum random circuit models that will be the main focus of this paper. We first review the equilibrium definition of generalized cluster models. Then, relying on the insights gained from Ref.~\cite{Lavasani2021},  we turn to the nonequilibrium scenario and formulate the random circuit models that realize their out-of-equilibrium counterparts. We argue that these models display various dynamical phases with SPT order and/or SSB, and we also construct a set of nonlocal string order parameters and SSB local order parameters, allowing us to determine the full phase diagram.

The family of generalized cluster models in a one-dimensional spin chain~\cite{generalized_cluster_models, Verresen17} is generated by Hamiltonians of the form
\begin{equation}
\label{eq:hamiltonians}
    H_\alpha = -\sum_n X_n\underbrace{Z_{n+1}\dots Z_{n+\alpha-1}}_{\alpha-1}X_{n+\alpha}, \quad \alpha \geq 1,
\end{equation}
where $X_n$, $Y_n$, and $Z_n$ denote the Pauli matrices at site $n$, and $\alpha$ is a positive integer parametrizing the members of the class. For a given $\alpha$, the model is symmetric under a set of $\alpha$ global symmetries,
\begin{equation}\label{eq:symmetries}
    G_1=\prod_k Z_{\alpha k +1},\, \dots ,\, G_{\alpha} = \prod_k Z_{\alpha k + \alpha}.
\end{equation}
Each of the symmetries $G_i$ is a product of $Z$ operators, distanced by $\alpha-1$ sites in the chain. 

All terms in the Hamiltonian $H_\alpha$, the so-called cluster operators $g^\alpha_n$, commute with each other, and thus the ground states are defined by the condition $g_n^\alpha\ket{\psi^\alpha_0} = \ket{\psi^\alpha_0}$ for every $n$. Such ground states realize different types of phases, with SSB and/or SPT order~\cite{generalized_cluster_models, Verresen17}. For example, $\alpha=1$ corresponds to the $\mathbb{Z}_2$ symmetric Ising chain, displaying SSB. The $\alpha=2$ case is the so-called cluster model that realizes an SPT phase, protected by $\mathbb{Z}_2\times\mathbb{Z}_2$ symmetry~\cite{Son_2011_spt_cluster, PhysRevA.84.022304_spt_cluster, zeng2016topological}. For $\alpha=3$, Eq.~\eqref{eq:hamiltonians} defines the XZZX cluster model with coexisting SSB and SPT orders~\cite{zeng2016topological, Verresen17}. In general, every odd  $\alpha$ value is characterized by $\mathbb{Z}_2$ symmetry breaking and $\mathbb{Z}_2^{\times (\alpha-1)}$ SPT order (except for $\alpha=1$), while even integers yield pure SPT order protected by a $\mathbb{Z}_2^{\times \alpha}$ symmetry.

\begin{figure}[t!]
    \centering
    \includegraphics[width=0.48\textwidth]{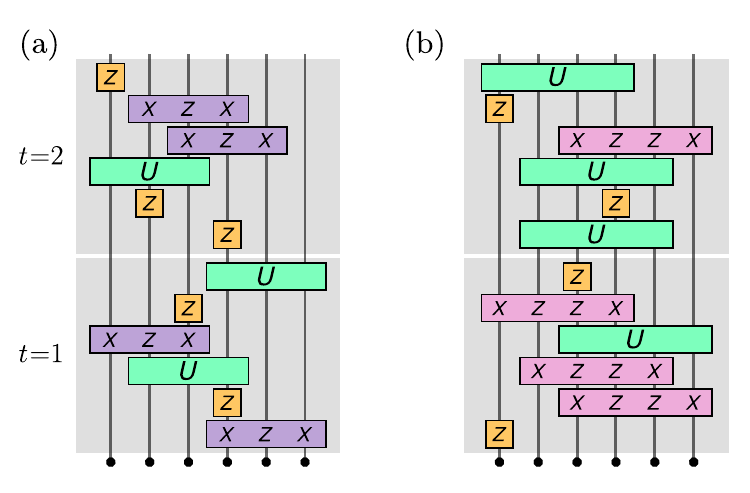}
    \caption{A particular realization of the cluster circuit model for (a) $\alpha=2$ and (b) $\alpha=3$, showing the first two time steps for system size $N=6$. The boxes labeled as $Z$, XZX and XZZX represent projective measurements, whereas the ones with label $U$ denote random unitary gates preserving the symmetries ~\eqref{eq:symmetries}.}
    \label{circuit_models}
\end{figure}

In Ref.~\cite{Lavasani2021}, Lavasani \textit{et al.} showed that the SPT phase of the XZX cluster model, $\alpha=2$, can be realized in a quantum random circuit by implementing properly designed measurements. Here we generalize this construction to induce by measurements the SPT phases that are realized by the generalized cluster models \eqref{eq:hamiltonians}. To this end, we construct a set of random circuit models in the following way. We consider a chain of $N$ qubits subject to open boundary conditions, with an initial state that can be an arbitrary eigenstate of the symmetry operators $G_1,\dots,G_\alpha$, e.g., the trivial product state $\ket{Z=1}^{\otimes N}$. At each step, we update the state by applying a sequence of three different operations:
\begin{enumerate}
    \item With probability $p_t$, we measure a cluster operator $X_iZ_{i+1}\dots Z_{i+\alpha-1}X_{i+\alpha}$, with $i\in\{1,\dots,N-\alpha\}$  chosen randomly.
    \item With probability $p_s$, we measure a single-qubit operator $Z_i$ on a random site $i$  of the chain.
    \item With probability $p_u=1-p_s-p_t$, a random unitary preserving the protecting symmetries~\eqref{eq:symmetries} and acting on $\alpha+1$ neighboring qubits is sampled and applied at a random position. The length of these unitaries, $\alpha+1$, was chosen as the shortest range such that the gates can create entanglement while preserving the symmetries of the model.
\end{enumerate}
The collection of $N$ such operations forms a single \textit{time step} in the time evolution of the system. Figure~\ref{circuit_models} illustrates this construction by showing the first two time steps in a particular realization of the circuit for $\alpha=2$ (left) and $\alpha=3$ (right).  We denote the \textit{circuit average} of a quantity $A$, measured in the steady state of the circuit, by $\overline{A}$. We note that in this setting, time average is equivalent to ensemble average; therefore, we can choose to perform the averaging over time steps and/or over circuit realizations.

This family of circuit models realizes three different phases for each $\alpha$, depending on the dominant operation. For large enough probability $p_u$, the unitary evolution dominates, resulting in a volume law phase with the entanglement entropy of subsystems scaling with their volume. In contrast, if measurements are applied at a sufficiently high rate, the stationary state is characterized by area law entanglement scaling. In particular, for large enough $p_s$, the $Z$ measurements tend to collapse the state to a trivial area law phase, whereas the measurements of cluster operators $X_iZ_{i+1}\dots Z_{i+\alpha-1}X_{i+\alpha}$ may induce  area law phases with symmetry-protected topological phases, and/or ordered phases with a local order parameter. For example, the case $\alpha=1$ has been studied in Ref.~\cite{Sang_2021} and was found to realize the out-of-equilibrium counterpart of the SSB order in the ground state of the Ising Hamiltonian for large enough $p_t$, dubbed a spin glass phase. For $\alpha=2$, Lavasani \textit{et al.} demonstrated the emergence of an SPT phase~\cite{Lavasani2021}.

Before investigating the phase diagram of the first few members of this family of random circuit models in detail, we comment on the order parameters that can distinguish different phases. One indicator that has been successfully applied in previous works is a topological entanglement entropy, readily accessible in numerical simulations for Clifford circuits, relying on the stabilizer formalism. However, in simulations of more general random circuits, for instance, in the presence of general Haar unitary gates $U$, as well as in possible experimental realizations, topological entanglement entropies are challenging to access. Therefore, we propose another way to distinguish the different phases by extending the concept of string order parameters, designed to detect equilibrium SPT phases~\cite{string_order,detection_spt_string_order}, to this out-of-equilibrium scenario. In general, for a state that is invariant under a protecting symmetry $G$ that can be expressed as the product of local unitary operators $\Sigma_i$, $G=\prod_i \Sigma_i$,  a string order parameter corresponding to boundary operators $O^{L/R}$ can be defined as follows~\cite{string_order,detection_spt_string_order},
\begin{equation}\label{eq:string}
    \mathcal{S}_\Sigma^{O^L, O^R} = \lim_{|j-k|\rightarrow\infty} \ev**{O^L(j)\left(\prod_{i=j+1}^{k-1}\Sigma_i\right)O^R(k)}{\psi_0}.
\end{equation}
These string order parameters allow us to differentiate topologically distinct states, by choosing the operators $O^{L/R}$ appropriately. Importantly, the bulk part of the string operator is constructed from the symmetry operators $\Sigma_i$. Therefore, for a symmetric state and for generic operators $O^{L/R}$, the string order parameter takes a nonzero value and varies smoothly as a function of the parameters of the Hamiltonian.  In order to distinguish topologically distinct phases, the boundary operators $O^{L/R}$ have to be chosen carefully, such that the emerging SPT order gives rise to selection rules that ensure the vanishing of a string order in a certain phase~\cite{detection_spt_string_order}. This method thus allows to detect topological phases through the exact relation $\mathcal{S}_\Sigma^{O^L, O^R}\equiv 0$. Choosing various pairs $O^{L/R}$, such that the resulting string orders $\mathcal{S}_\Sigma^{O^L, O^R}$ vanish in different phases, grants access to the full phase diagram. Convenient string order parameters for the generalized cluster models are given by
\begin{equation}
\label{eq:triv_string}
    \mathcal{S}_{Z}^{\mathds{1}, \mathds{1}}(\alpha) = \lim_{|j-k|\rightarrow\infty} \expval{\prod_{i=j + 1}^{k}Z_{\alpha i}},
\end{equation}
vanishing in the ground state of the cluster Hamiltonian, as well as
\begin{equation}
\small
\label{eq:spt_string}
    \mathcal{S}_{Z}^{X, X}(\alpha) = \lim_{|j-k|\rightarrow\infty} \expval{X_{\alpha j}\left(\prod_{i=j + 1}^{k}Z_{\alpha i-\alpha+1}\dots Z_{\alpha i-1}\right)X_{\alpha k}},
\end{equation}
yielding zero for a trivial $Z$ product state.

Besides the SPT order, for odd $\alpha$ values, the ground state of Hamiltonian~\eqref{eq:hamiltonians} displays SSB.  This type of order can be detected through the connected correlators of a symmetry-breaking local order parameter, defined as
\begin{equation}
\label{eq:local_order_parameter}
    M(\alpha)=X_i\underbrace{(Y_{i+1}X_{i+2})\dots(Y_{i+\alpha-2}X_{i+\alpha-1})}_{(\alpha+1)/2}.
\end{equation}
We note that for $\alpha=1$, $M(\alpha)$ reduces to the Ising order parameter $X_i$.

While the various order parameters defined in Eqs~\eqref{eq:triv_string}, ~\eqref{eq:spt_string} and ~\eqref{eq:local_order_parameter} are well-suited for determining the phase diagram of the generalized cluster models in equilibrium, they are not directly applicable for the random circuit scenario. The reason for this is that for different realizations of the disordered circuit, the sign of string order parameters and correlation functions fluctuates randomly, yielding a vanishing circuit average in all nontrivial regions of the parameter space. Similarly, the time average of all of these quantities vanishes for any fixed circuit realizations. This property can be understood by noting that all the measured operators have eigenvalues $\pm1$, resulting in a randomly changing sign during the time evolution because of the probabilistic measurement outcomes and the repeatedly applied random unitary gates. 

The vanishing of these circuit averages is also intimately related to the nature of the measurement-induced entanglement transition. This dynamical phase transition is unconventional in the sense that it relies on the properties of individual quantum trajectories instead of the disorder-averaged quantum state, and it can only be detected through quantities that are nonlinear in the density operator of the system, e.g., through entanglement entropies~\cite{review_random_circuits,Li_2019}. In contrast, the circuit average of the density matrix is a trivial infinite temperature density matrix. Therefore, the rich entanglement structure of individual quantum trajectories remains hidden at the level of the average density matrix, and, consequently, at the level of disorder-averaged operator averages, such as the string order parameters or connected correlators considered above.

This difficulty can be overcome by modifying the proposed order parameters in such a way that they become nonlinear in the density matrix of the system. This can be easily achieved by considering the time and/or circuit average of the \textit{absolute value} of the string order parameters, $\overline{|\mathcal{S}|}$, with a similar idea applied to the correlator of the local order parameters $M(\alpha)$. We test these proposed order parameters below by benchmarking them against the behavior of the topological entanglement entropy for various circuit models and demonstrate that they capture the full phase diagram correctly. Nevertheless, while these string order parameters are readily accessible in tensor network simulations, it is important to remark that they are still subject to the so-called postselection problem, i.e., determining them experimentally requires preparing many copies of the steady state for each set of measurement outcomes. Due to the probabilistic nature of measurements, it is exponentially unlikely to find the same measurement outcomes in two realizations of the same circuit,  hindering the observation of measurement-induced phase transitions, despite a number of possible solutions proposed theoretically in recent years~\cite{Noel_2022, Dehghani_2023, Li_2023}. 

Before turning to the numerical simulations, we briefly comment on the special case of $p_u=0$, resulting in models without unitary evolution. We find that these circuits, consisting only of projective measurements, share certain universal properties for any value of $\alpha$. In particular, they all display a phase transition between the trivial and the SPT and/or SSB area law phase for $p_s=1/2$, i.e., where the rates of both types of measurements are equal. A proof of this statement, relying on a duality argument, is presented in Appendix~\ref{appendix_A}.

\section{SPT phase in the XZX cluster circuit model}\label{sec:xzx}

In this section, we revisit the SPT phase of the XZX cluster circuit model, by applying the framework presented in Sec.~\ref{sec:framework}. We first focus on Clifford circuits in Sec.~\ref{subsec:clifford}, already studied in Ref.~\cite{Lavasani2021}. In this special set of circuits, both topological entanglement entropies and string order parameters are readily accessible by relying on the stabilizer formalism. We benchmark the string order parameters~\eqref{eq:triv_string} and~\eqref{eq:spt_string} by using them to detect the phase transitions and comparing the phase diagram to the one obtained from entanglement entropies. We then turn to more general Haar random circuits in Sec.~\ref{subsec:haar}, extending the class of random circuits compared to Ref.~\cite{Lavasani2021}, and testing the stability of the phase diagram against allowing a wider set of unitary gates during the time evolution. Here the stabilizer formalism is no longer applicable. Instead, we rely on efficient tensor network methods to simulate the dynamics and extract the string order parameters~\eqref{eq:triv_string} and~\eqref{eq:spt_string} in order to determine the full phase diagram. In accordance with general expectations, we find that the area law phases shrink slightly upon  extending the class of unitary gates; nevertheless, the phase diagram remains qualitatively very similar  to the one obtained for Clifford circuits. These results provide evidence for the robustness of area law phases in different random circuit models.

\subsection{Time evolution with Clifford unitary gates}\label{subsec:clifford}

\begin{figure}[t!]
    \centering
    \includegraphics{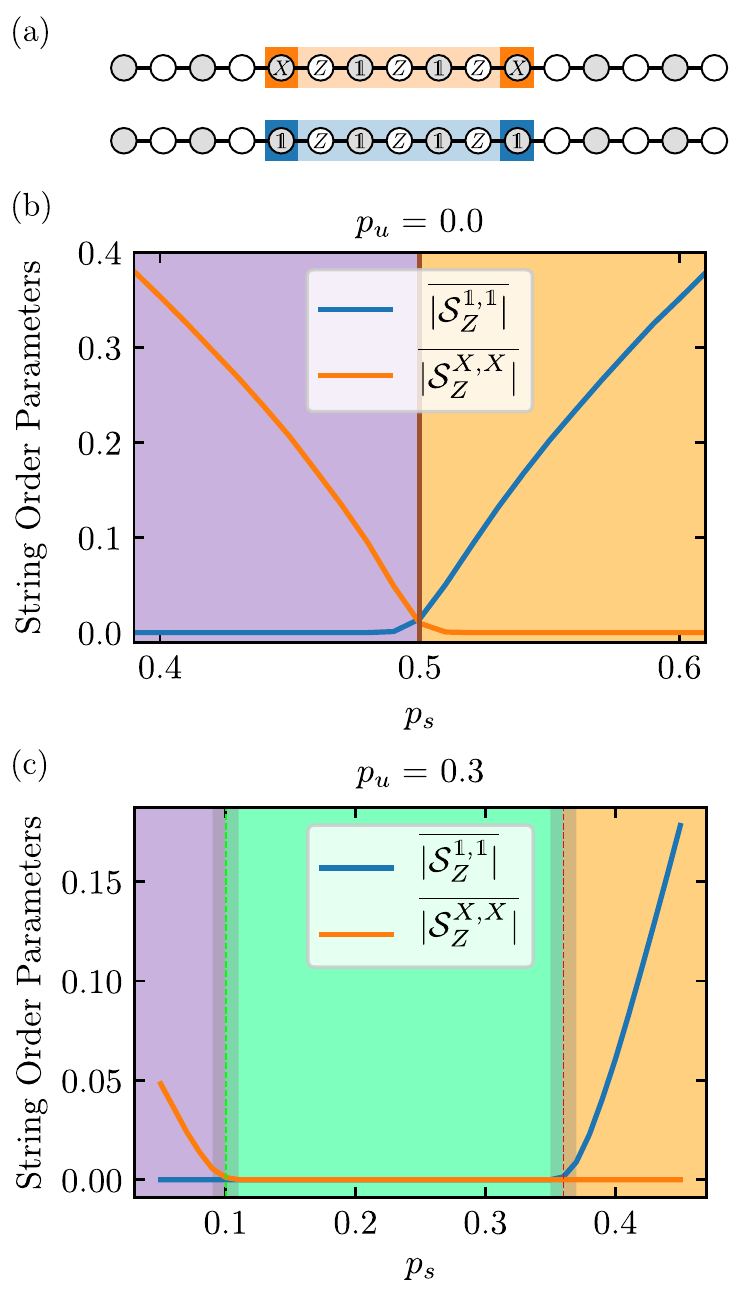}
    \caption{String order parameters in a Clifford circuit for $\alpha=2$. (a) String operators~\eqref{eq:spt_string} and~\eqref{eq:triv_string} in a finite lattice, characterized by the same bulk operator (light shading), and distinguished by the different boundary operators (dark shading). (b) String order parameters for measurement-only dynamics $p_u=0$, shown as a function of the probability of single-qubit measurement $p_s$ in the vicinity of  the self-dual point $p_s=0.5$. String order parameters distinguish two area law phases, one with SPT order (purple) and a trivial one (orange). The vertical line indicates the exact critical point of the phase transition, $p_s=0.5$. (c) String order parameters in the presence of a finite rate of unitary gates, $p_u=0.3$, shown as a function of $p_s$ across the phase boundaries. Unitary gates are Clifford unitaries respecting  the protecting $\mathbb{Z}_2\times\mathbb{Z}_2$ symmetry. String order parameters distinguish an SPT (purple) and a trivial (orange) area law phase, separated by a volume law region (green). Numerically determined phase boundaries are indicated by vertical dashed lines. In (b) and (c) we used system size $N=1024$.}
    \label{fig:XZX_Clifford}
\end{figure}

In this section, we reproduce the phase diagram of Ref. ~\cite{Lavasani2021}, by relying on the string order parameters instead of a topological entanglement entropy. The special structure of  Clifford unitary gates allows us to efficiently simulate large system sizes, up to $N=1024$ qubits, by applying the stabilizer formalism. This method relies on representing the wave function of the system for a given circuit realization as the eigenstate of $N$ linearly independent (under multiplication) commuting Pauli strings, the so-called stabilizers. Both Clifford unitaries and projective measurements preserve this structure by mapping the stabilizers to another set of $N$ independent commuting Pauli strings, allowing to simulate the time evolution efficiently~\cite{AaronsonGottesman}. The circuit averages are then obtained by performing the averaging over the stationary states of  $10^3$ random circuits. For each of these circuits, the qubits are initialized in the state $\ket{Z=-1}^{\otimes N}$, and let to evolve for $2N$ time steps (corresponding to $2N^2$ operations) to reach the steady state. Then, an additional time averaging is performed by evolving the system for another $10^3$ time steps, calculating the desired string expectation values after each of them, and taking the average of the obtained values.

For the finite systems considered here, we choose string operators of length $N/2-1$, located in the middle of the chain, as depicted in Fig.~\ref{fig:XZX_Clifford}a. We also make use of sublattice symmetry in the following way. The string illustrated in Fig.~\ref{fig:XZX_Clifford}a, displaced by one site to the right,  will lead to the same circuit average. Therefore, we can improve the convergence of disorder averaging by averaging our results for the original and shifted strings, for both string order parameters~\eqref{eq:triv_string} and~\eqref{eq:spt_string}.

Before turning to the time evolution in the presence of unitary gates, we first comment on the special case of measurement-only dynamics, $p_u=0$, a  line in parameter space that is the same for Clifford and Haar random circuits. As shown in Fig.~\ref{fig:XZX_Clifford}b, the string order parameters allow to distinguish the two different area law phases in the model. The SPT phase (purple) is characterized by
$\overline{|\mathcal{S}_Z^{X,X}|}>0$ and $\overline{|\mathcal{S}_Z^{\mathds{1},\mathds{1}}|}=0$, while the trivial phase is signaled by $\overline{|\mathcal{S}_Z^{X,X}|}=0$ and $\overline{|\mathcal{S}_Z^{\mathds{1},\mathds{1}}|}>0.$ Based on duality arguments, this transition happens exactly at $p_s=1/2$, in good agreement with our numerical results. We note that at the critical point $p_s=1/2$, both string order parameters should vanish in the thermodynamic limit. Due to finite-size effects, we find a small finite value instead for the fixed system size $N=1024$ shown in Fig.~\ref{fig:XZX_Clifford}b.

Next, we turn to the case involving random Clifford unitary gates, preserving the $\mathbb{Z}_2\cross\mathbb{Z}_2$ symmetry~\eqref{eq:symmetries}. Figure~\ref{fig:XZX_Clifford}c shows the behavior of the string order parameters as a function of the single-qubit measurement probability $p_s$ for a fixed rate of unitary evolution, $p_u=0.3$. We find that the string operators are still well suited for reconstructing the phase diagram. Besides the area law  SPT (purple) and trivial (orange) phases already seen for measurement-only dynamics, the volume law phase (green) is clearly distinguished by the vanishing of both string order parameters.

Our results demonstrate that string order parameters provide an accessible alternative way to determine the phase boundaries. To account for finite-size effects,  we calculate the string order parameters $\overline{|\mathcal{S}|}(N)$ for various system sizes $N$. At the critical point, this quantity is expected to decay towards zero as a power law in $N$, allowing us to implement an extrapolation to the thermodynamic limit, as discussed in Appendix~\ref{appendix_C}. The phase diagram determined with this method is shown in Fig.~\ref{fig:XZX_phase_diagrams}a, and is in perfect agreement with the results of Ref.~\cite{Lavasani2021}.

\subsection{Phase diagram for Haar random unitary gates}\label{subsec:haar}

\begin{figure}[b!]
    \centering
    \includegraphics{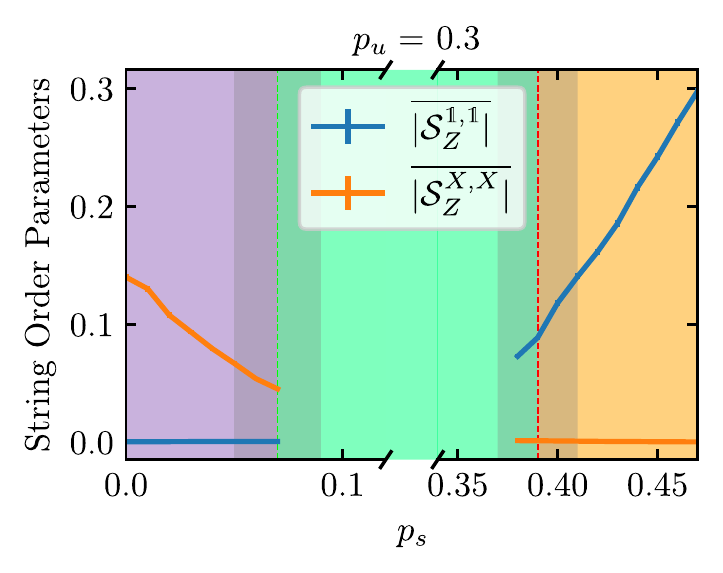}
    \caption{String order parameters  in the $\alpha=2$ cluster circuit model with Haar random unitary gates, plotted as a function of single spin measurement probability $p_s$ across the phase boundaries, for fixed $p_u=0.3$. These order parameters distinguish an SPT (purple) and a trivial (orange) area law phase, separated by a volume law phase (green). The vertical dashed lines show the numerically determined phase boundaries, with the grey shadow indicating the uncertainty. We used system size $N=128$ and an adaptive bond dimension to ensure convergence for all data points, with a maximal value $\chi_{\text{max}}=512$ used in the points close to the phase transition. Data are not shown for the volume law phase (green), where  MPS simulations are not efficient and did not convergence for  available bond dimensions.}
    \label{fig:XZX_MPS_pu0.3_128}
\end{figure}

In this section, we check the stability of the phase diagram of the circuit model with respect to a broader class of unitary gates. To this end, we consider the time evolution in the presence of generic random Haar unitaries, with the only restriction that the gates still preserve the protecting symmetries~\eqref{eq:symmetries}. Thereby, we considerably extend the set of allowed gates in the circuit compared to the special Clifford gates considered earlier. In principle, relaxing this constraint on the structure of gates could lead to a more efficient entanglement generation during the time evolution, and might lead to a much broader region of volume law phase in parameter space. Therefore, we examine the sensitivity of phase boundaries towards such an extension of allowed quantum circuits. As we will demonstrate below, we can still identify two distinct area law entangled phases, displaying SPT and trivial order, respectively, in analogy with the Clifford case. 

For general Haar random gates, an efficient simulation of the circuit relying on the stabilizer formalism is no longer possible. Instead, we represent the wave function using matrix product states (MPS)~\cite{DMRG_MPS, tenpy}, and implement the time-evolving block decimation (TEBD) algorithm to calculate the dynamics~\cite{TEBD_algorithm}. This method is well-suited for studying the SPT and trivial area law phases~\footnote{The existence of an efficient MPS representation of the area law state is only proved for the case when the Rényi entropy $S_{n<1}$ is bounded by $\log N$. In other cases, it is, in general undetermined whether the MPS approximates the state properly, see Appendix~\ref{appendix_C} for more details.}; however, the MPS representation with finite bond dimension breaks down in the volume law phase. Nevertheless, we can still find indications of the volume law phase through the logarithmically increasing half-chain entanglement entropy as a function of the maximal allowed bond dimension, as discussed in Appendix~\ref{appendix_B}. We also note that measuring the topological entanglement entropy is much more demanding both in our MPS simulations and in an experimental setup. Therefore, here we solely rely on more accessible string order parameters to determine the phase diagram.

In our MPS simulations, we consider spin chains with up to $N=128$ qubits.  To ensure the validity of the results, we examine the convergence of the entanglement entropy with increasing bond dimension, and we keep track of the symmetries ~\eqref{eq:symmetries} during the time evolution, to ensure that they are not broken due to the truncation in the TEBD algorithm. The required bond dimension is found to increase as the critical lines are approached, and we run simulations with up to $\chi_{\text{max}}=512$. We discuss the convergence with bond dimension in detail in Appendix~\ref{appendix_B}. We note that in the SPT phase it is important to choose a bond dimension $\chi_{\text{max}}$ that is a power of 2, to minimize the probability of breaking the symmetries by truncating degenerate Schmidt values, leading to changes of steady-state properties (see Appendix~\ref{appendix_D}). We obtain the circuit averages by generating $10^2$ random circuits, with the qubits initialized in the state $\ket{Z=-1}^{\otimes N}$. We also perform an additional time averaging over $2\times10^2$ time steps after the system has reached the steady state.

\begin{figure}[t!]
    \centering
    \includegraphics[width = 0.47\textwidth]{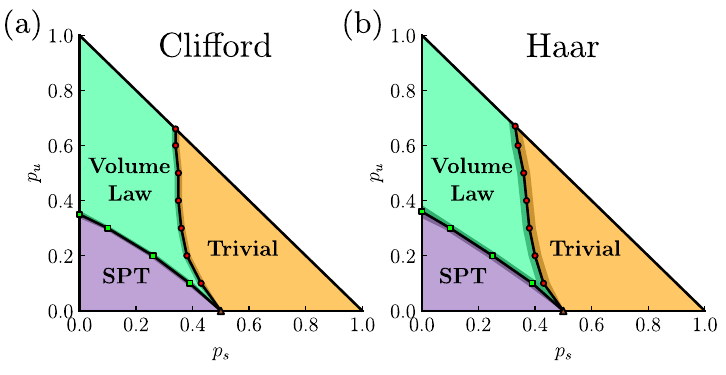}
    \caption{Phase diagram determined from string order parameters, (a) with random Clifford unitary gates, (b) with random Haar unitaries, preserving the $\mathbb{Z}_2\cross\mathbb{Z}_2$ symmetry. Numerical uncertainty is indicated by grey shades. Both phase diagrams display an SPT and a trivial area law phase, separated by a volume law region. Clifford results are consistent with the phase diagram obtained in Ref.~\cite{Lavasani2021}, relying on entanglement entropies.}
    \label{fig:XZX_phase_diagrams}
\end{figure}

The string order parameters are shown as a function of $p_s$ in Fig.~\ref{fig:XZX_MPS_pu0.3_128}, for a fixed probability of unitary gates $p_u=0.3$ and system size $N=128$. All data points shown here are converged with respect to bond dimension. For a low enough rate of $Z$ measurements $p_s$, our simulation yields string order parameters $\overline{|\mathcal{S}_Z^{X,X}|}>0$ and $\overline{|\mathcal{S}_Z^{\mathds{1},\mathds{1}}|}=0$, consistent  with the presence of an SPT phase also for random Haar unitary gates (purple region). Similarly, $\overline{|\mathcal{S}_Z^{X,X}|}=0$ and $\overline{|\mathcal{S}_Z^{\mathds{1},\mathds{1}}|}>0$ indicate a trivial area law phase (orange) for large enough $p_s$. We confirm these predictions by carefully examining the finite-size scaling of the string order parameters and extrapolating them to the thermodynamic limit (see Appendix~\ref{appendix_C}). This procedure leads to the phase boundaries shown by vertical dashed lines, with the two area law regions  separated by a volume law phase (green), characterized by the vanishing of both string order parameters.  While the MPS representation with finite bond dimension  breaks down  in this region, for the available system sizes and bond dimensions we observed that the scaling of the half-chain entanglement entropy is consistent with a volume law behavior, providing additional support for the conclusions drawn from the string order parameters.  We note that the area law regions in the Haar case shrink slightly compared  to the Clifford circuit (cf. to Fig.~\ref{fig:XZX_Clifford}c), in accordance with the expectation that extending the set of allowed unitary gates drives the system to a more entangled state.

We numerically estimate the full phase diagram for the Haar random circuit by performing a similar extrapolation to the thermodynamic limit, varying the probabilities $p_s$ and $p_u$ (see Appendix~\ref{appendix_C}). To this end, we extracted the value of the string order parameters for various system sizes at each point in parameter space,  and we carefully verified the convergence with bond dimension for all system sizes used in this extrapolation procedure. We considered the area law phase up to the vicinity of criticality, where the required bond dimension is expected to scale as $\xi_{\text{max}}\propto N$. The results are shown in Fig.~\ref{fig:XZX_phase_diagrams}b, compared to the phase diagram of the Clifford circuit, Fig.~\ref{fig:XZX_phase_diagrams}a. The estimated numerical uncertainty of the phase boundaries is indicated by grey shading, with the Haar circuit results subject to a larger error due to the system size limitations of the MPS simulations.

As a final remark, we note that MPS simulations grant us access to the scaling exponents of the phase transitions, and allow to address the question of whether the Haar and Clifford circuits belong to the same universality class.  For completeness, we perform a finite-size scaling collapse for the string order parameters, yielding  two critical exponents $\nu_s$ and $\eta$, defined by the scaling ansatz
\begin{equation*}
\label{scaling_ansatz}
    \mathcal{S}_\Sigma^{O^L, O^R}(p, N) = N^{-\eta} F\left((p-p_c)N^{1/\nu_s}\right).
\end{equation*}
Here $p_c$ denotes the critical point, and $F$ is the scaling function.
Our best numerical estimates are consistent with the same correlation length exponent  $\nu_s$ for Clifford and Haar circuits, while the exponent $\eta$ is model dependent. We can also determine a critical correlation length exponent $\nu$ from the scaling of the entanglement entropy for the phase transition between the trivial and volume law phase. Our numerical findings show that both $\nu_s$ and $\nu$ are consistent with the percolation value $\nu_{\rm perc}=4/3$ up to numerical precision.  We note, however, that these results are subject to considerable numerical uncertainty, warranting further large-scale simulations, that we leave for future work. For a more detailed discussion of our results on finite-size scaling and critical exponents, please visit Appendix~\ref{appendix_E}.

\section{Coexistence of SPT and SSB}\label{sec:xzzx}

Having validated our approach in the previous section, we now turn to other members of the family of generalized cluster models. In this section, we demonstrate that a coexisting SPT order and SSB can be realized in the area law phase of the circuit models for odd $\alpha$ values. To this end, here we focus on the case with $\alpha=3$, realizing the XZZX circuit model shown in Fig.~\ref{circuit_models}b. The ground state of the corresponding generalized cluster model Hamiltonian ~\eqref{eq:hamiltonians} shows both SPT and SSB orders. As we demonstrate below, the out-of-equilibrium version displays two area law phases: one characterized by coexisting SPT and SSB orders and a trivial phase, separated by a volume law entangled phase. In this section, we restrict our attention to Clifford circuits, with the unitary gates preserving the symmetries ~\eqref{eq:symmetries}, allowing us to reach large system sizes by applying the stabilizer formalism. The measurement-only limit of this model has been recently studied in Ref.~\cite{Klocke_2022}.

\begin{figure}[t!]
    \centering
    \includegraphics{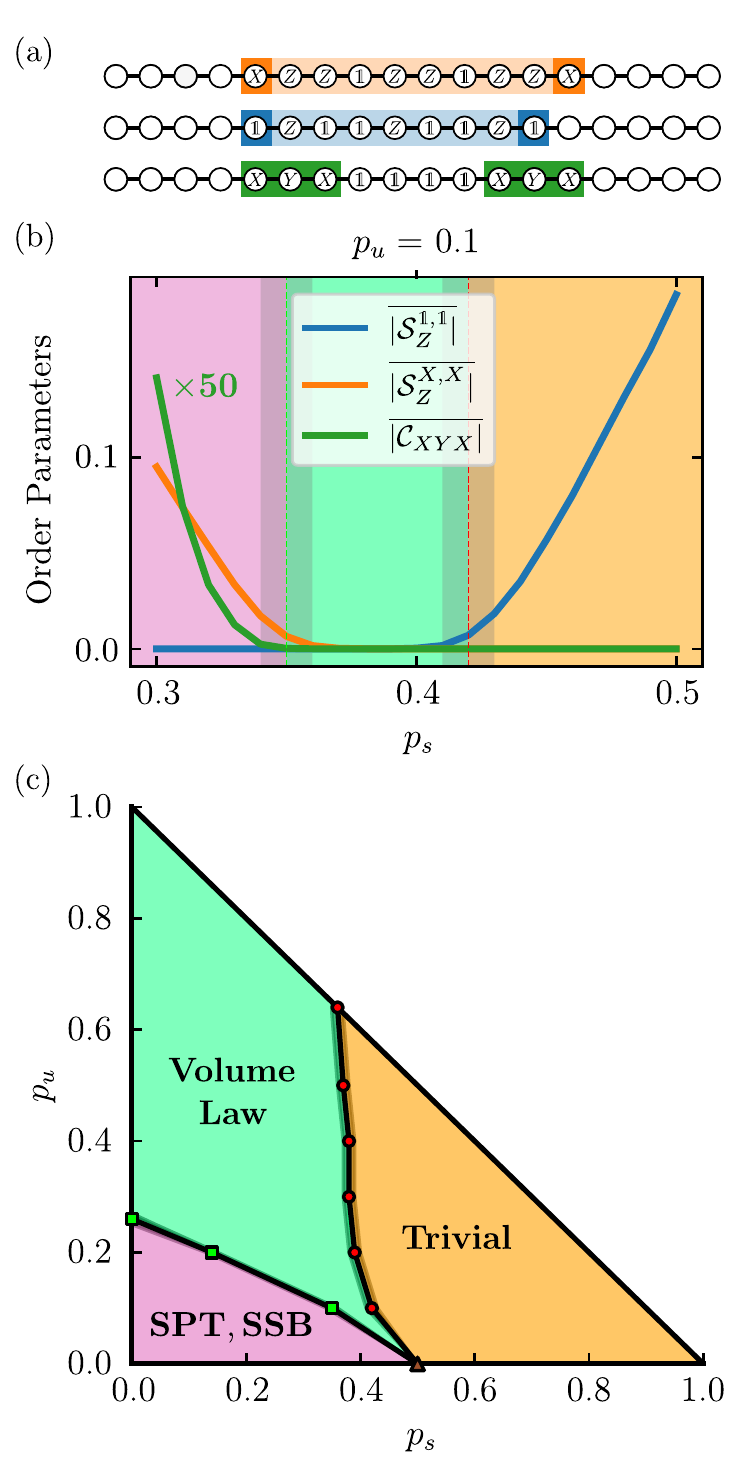}
    \caption{Phases realized by the XZZX ($\alpha=3$) circuit model. (a) String operators and  connected correlator of the local order parameter in a finite lattice. The two string operators (top and middle) are characterized by a nontrivial bulk operator (light shading), and boundary operators (dark shading), whereas the local order parameter (bottom) is measured at two distant positions (dark shading) to capture SSB. (b) Order parameters across the two phase transitions versus the probability of single-qubit measurement, $p_s$, at a fixed rate of unitary gates, $p_u=0.1$, for system size $N=768$. The two string operators and the correlator (multiplied by 50 for better visibility) reveal an area law phase with coexisting SPT and SSB order (magenta), as well as a trivial area law phase (orange), separated by a volume law phase (green). Vertical lines indicate the numerically determined phase boundaries.   (c) Full phase diagram of the XZZX circuit model obtained via string order parameters.}
    \label{fig:XZZX_model}
\end{figure}

To differentiate the phases realized by this model, we use both the string order parameters of Eqs.~\eqref{eq:triv_string} and ~\eqref{eq:spt_string},  and the local order parameter of Eq.~\eqref{eq:local_order_parameter},  yielding $M_i=X_{i}Y_{i+1}X_{i+2}$ in this case. The SSB ordering can be detected through the correlators of this local order parameter,
\begin{equation}\label{eq:correlator}
    \mathcal{C}_M = \lim_{|j-k|\rightarrow\infty}\expval{M_jM_k}.
\end{equation}
The finite-size versions of the string order parameters, as well as the correlation function of Eq.~\eqref{eq:correlator} used for this model, are illustrated in Fig.~\ref{fig:XZZX_model}a. The initial state of the qubits, as well as the procedure for obtaining the circuit averages of various quantities, is the same as the one applied for the XZX model. 

We show the behavior of the circuit-averaged string order parameters and the correlator of the local order parameter in Fig.~\ref{fig:XZZX_model}b, for a fixed rate of unitaries $p_u=0.1$, using system size $N=768$. For a small rate of $Z$ measurements $p_s$, the string order parameter $\mathcal{S}_{Z}^{\mathds{1}, \mathds{1}}$ vanishes, while the other string operator $\mathcal{S}_{Z}^{X,X}$ and the correlator $\mathcal{C}_{XYX}$  both take finite values. This indicates an area law phase characterized by the coexistence of SPT and SSB orders (magenta). We note that up to numerical precision, the SPT order and SSB vanish at the same critical point; thus we do not find an area law phase displaying solely SPT or SSB order in this model. As shown in Ref.~\cite{Klocke_2022} in the limit of measurement-only dynamics, an SPT phase without SSB order can be generated by adding symmetry-breaking operations, such as $Y$ measurements, to the circuit. For large single-qubit measurement probability $p_s$, we find a trivial area law phase (orange) with vanishing string order parameter $\mathcal{S}_{Z}^{X,X}$ and correlator $\mathcal{C}_{XYX}$, but a finite value for $\mathcal{S}_{Z}^{\mathds{1}, \mathds{1}}$. As before, the two area law regions are separated by a volume law phase (green), where all order parameters become zero.

The full phase diagram of this model, extracted from the extrapolation to the thermodynamic limit of the three types of order parameters (as detailed in Appendix~\ref{appendix_C}), is shown in Fig.~\ref{fig:XZZX_model}c. By comparing it to the phase diagram of the XZX model, Fig.~\ref{fig:XZX_phase_diagrams}, we observe that the volume law phase becomes more extended in the XZZX model, and the region with SPT and SSB orders decreases in size with respect to the SPT phase of the XZX circuit. This effect stems from the longer range of the random unitary gates in the XZZX circuits, leading to more efficient entanglement generation. Similarly to the XZX model, the numerical results, as well as the analytical arguments presented in Appendix~\ref{appendix_A}, suggest that the point $p_s=p_c=1/2$ and $p_u=0$ is a tricritical point.

\section{Conclusion}\label{sec:conclusion}

We have studied the interplay of symmetry-protected topological phases and measurement-induced entanglement transition by introducing a class of quantum random circuit models, consisting of projective measurements and random unitary gates respecting a set of global symmetries. We showed that the circuits in this family display measurement-induced phase transitions between a thermal volume law phase and different nonthermal area law stationary states, and realize the out-of-equilibrium version of all generalized cluster states in their area law phase. Motivated by the string operators used to detect SPT order in equilibrium settings, we have constructed a set of nonequilibrium string order parameters, well suited for revealing SPT order in this class of circuit models, and readily accessible to numerical simulations. We benchmarked our framework by studying the string order in the XZX circuit model and comparing it to the behavior of the topological entanglement entropy in the special case of Clifford unitary gates. We then tested the stability of the phase diagram by relaxing the strong constraint on the structure of unitary gates, and allowing for a wider set of symmetry-preserving Haar random unitaries. Relying on MPS simulations, we studied the finite-size scaling  of the string order parameters, allowing us to distinguish two area law phases and a volume law phase and to determine the phase boundaries.  Our results confirm the presence of SPT order in an extended region in parameter space and provide additional evidence that the rich structure observed in Clifford circuits also appears in more generic quantum circuit models. Finally, we demonstrated in the example of the XZZX circuit model that the out-of-equilibrium generalized cluster states can host simultaneous SPT order and SSB, similarly to their equilibrium Hamiltonian counterparts.

Our results pave the way to study topological phases in a wider range of quantum circuit models. In particular, as we demonstrated in this work, the out-of-equilibrium string order parameters are accessible in MPS simulations, allowing us to study the phase diagram of generic Haar random circuits. One of the interesting open questions in this direction concerns the universality class of various entanglement transitions. We demonstrated that the MPS framework developed in this paper opens a way to study the universal behavior of Measurement-Induced Phase Transitions by examining the scaling of various order parameters and entanglement entropies from the area law side of the transition. We leave a more detailed analysis of these scaling properties for future work. Furthermore, it would be interesting to address how much information can be obtained about the volume law phase by studying the scaling of order parameters with bond dimension. Another interesting direction is characterizing all the possible ordered phases that can arise in the area law stationary states of random circuit models. The family of random circuits considered in this paper provides one example of a class of ordered phases; however, the construction could be extended to give a recipe for realizing other types of phases, such as true topological order in higher dimensional circuits. 

\textbf{Acknowledgements}.
This work was supported by the European Research Council (ERC) under the European Union’s Horizon 2020 research and innovation programme, grant agreement Nos. 851161 and 771537. F.P. acknowledges the support of the Deutsche Forschungsgemeinschaft (DFG, German Research Foundation) under Germany’s Excellence
Strategy EXC-2111-390814868.  F.P.’s research is part of
the Munich Quantum Valley, which is supported by the
Bavarian state government with funds from the Hightech
Agenda Bayern Plus. I.L. acknowledges support from the Gordon and Betty Moore Foundation through Grant GBMF8690 to UCSB and from the National Science Foundation under Grant No. NSF PHY-1748958.

\textbf{Data and materials availability.} Data analysis and simulation codes are available on Zenodo upon reasonable request~\cite{Zenodo}.

\begin{appendix}

\section{Phase transition in measurement-only random circuits}
\label{appendix_A}
\setcounter{equation}{0}
\renewcommand{\theequation}{A\arabic{equation}}

In this appendix we focus on random circuits in the measurement-only regime, $p_u=0$, displaying a single phase transition between two topologically distinct area law phases. Here we prove via a duality argument that the critical point is located at $p_s=1/2$ for any value of $\alpha$. This critical point is characterized by a logarithmic scaling of the entanglement entropy as argued in~\cite{EPT_only_measurements}. We note that our reasoning generalizes the proof from Ref.~\cite{Lavasani2021}, finding the critical point of the measurement-only XZX circuit cluster model, $\alpha=2$, by extending it to all values of $\alpha$.

Consider the circuit model for $\alpha>0$ and periodic boundary conditions, with initial state $\ket{\psi_0} = \ket{0, \dots, 0}$, described by the generating set of stabilizers $\{Z_1, Z_2, \dots, Z_{N-\alpha}, G_1, \dots, G_\alpha\}$. Here we make the symmetries of the model explicit, as stabilizers that remain unaltered throughout the evolution of the circuit. We denote the cluster operators by $g_i = X_{i-\alpha/2}Z_{i-\alpha/2+1}\dots Z_{i+\alpha/2-1}X_{i+\alpha/2}$, with $i\in \{1,\dots, N\}$ for $\alpha$ even and $i\in \{1/2, 3/2, \dots, N-1/2\}$ for $\alpha$ odd (note that all sums in indices are understood to be modulo $N$). At each step of the circuit we measure $Z_i$ with probability $p_s$ and $g_i$ with probability $1-p_s$. From the Gottesman-Knill theorem~\cite{measurementsGottesman,AaronsonGottesman}, a stabilizer at any time step of the evolution can be written as a product of $Z$ and $g$ operators.

\begin{figure}[t!]
    \centering
    \includegraphics[width=0.48\textwidth]{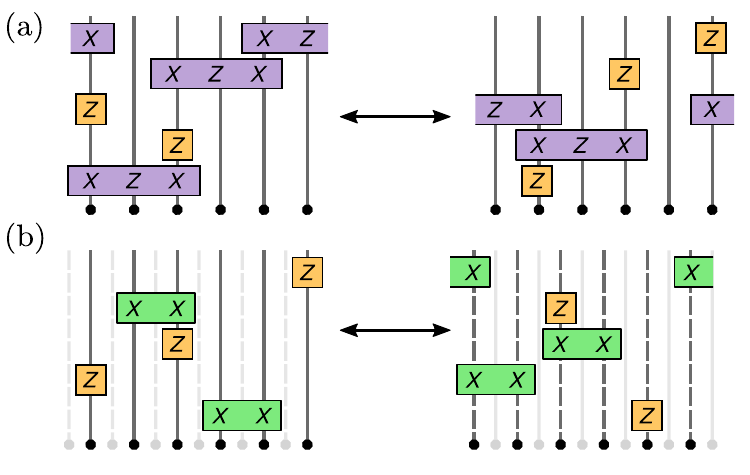}
    \caption{Duality map between circuit realizations for (a) $\alpha = 2$ and (b) $\alpha = 1$. For $\alpha$ even the mapping connects circuits defined on the same lattice, whereas for $\alpha$  odd we introduce a dual space. In both cases, the initial state of the original circuit is given by the stabilizers $\{Z_1, Z_2, \dots, Z_{N-\alpha}, G_1, \dots, G_\alpha\}$, and the initial state of the dual circuit is given by stabilizers $\{g_1,\dots,g_{N-\alpha}, G_1,\dots,G_\alpha\}$.}
    \label{fig:duality}
\end{figure}

For each realization of the circuit model with probability $p_s$ of $Z$ measurement, we construct a dual version with probability $1-p_s$ in the following way. For $\alpha$ even, the dual circuit will be defined on the same lattice as the original one, whereas for $\alpha$ odd we introduce a dual lattice, with lattice sites indexed by half integers $i\in \{1/2, 3/2, \dots, N-1/2\}$, and perform a mapping to this dual space. First, we set the initial state of the dual circuit to be the one fixed by the stabilizers $\{g_1,\dots,g_{N-\alpha}, G_1,\dots,G_\alpha\}$. Then, we substitute every $Z_i$ measurement by a $g_i$ measurement, and vice versa.  Fig.~\ref{fig:duality} shows the duality transformation for a realization of the circuit at $\alpha=2$ and $\alpha=1$. The evolved state of the dual circuit is closely related to the state of the original circuit, as the following Lemma shows.

\
\\
\textbf{Lemma}. Let 
\begin{equation}
    s = \left(\prod_{i\in I} Z_i\right)\left(\prod_{j\in J} g_j\right),
\end{equation}
be a stabilizer of the state of the original circuit at the updating step $M$, where $I\subset \{1,\dots, N\}$ and $J\subset\{1,\dots,N\}$ for $\alpha$ even and $J\subset\{1/2,\dots,N/2-1\}$ for $\alpha$ odd. Then, the state of the dual circuit at updating step $M$ is stabilized by the operator
\begin{equation}
    \tilde{s} = \left(\prod_{i\in I} g_i\right)\left(\prod_{j\in J} Z_j\right).
\end{equation}
\\
\\
\textit{Proof}. We prove the lemma by induction. By construction, it is true for the initial state. Suppose it is true at updating step $M$, and now let us apply a new measurement. If the new measurement commutes with all stabilizers, then the state is unchanged, so the claim is true. Let us consider the case where the measured operator does not commute with all the stabilizers. For concreteness, suppose that we measure the operator $Z_i$ in the original circuit (and thus $g_i$ in the dual one). The measurement anticommutes with all the stabilizers that, when expressed as a product of $g$ and $Z$ operators, contain either $g_{i+\alpha/2}$ or $g_{i-\alpha/2}$ (but not both). We denote this set of stabilizers by $\{s_1,\dots,s_n\}$. When $Z_i$ is measured, this set is updated to $\{Z_i, s_1s_2, \dots, s_1s_n\}$, by virtue of the Gottesman-Knill theorem. In the dual circuit, $g_i$ anticommutes with $Z_{i+\alpha/2}$ and $Z_{i-\alpha/2}$, so all the stabilizers commuting with these are $\{\tilde{s}_1, \dots, \tilde{s}_n\}$ due to the induction hypothesis. The updated state is stabilized by the operators $\{g_i, \tilde{s}_1\tilde{s}_2,\dots, \tilde{s}_1\tilde{s}_n\}$, so for the modified stabilizers the claim is still true ($s_1s_j \rightarrow \tilde{s}_1\tilde{s}_j$). The same argument holds for the case where the measured stabilizer is $g_i$. $\square$

To prove that the phase transition is located at $p_s=1/2$, we apply a duality argument. Below we show that if the state of one of the circuits has area law entanglement at some time step, then so does its dual counterpart. Therefore, the critical point with logarithmic entanglement scaling must coincide with the self-dual point $p_s=1/2$.  Within the stabilizer formalism~\cite{AaronsonGottesman}, the entanglement entropy of a stabilizer state $\ket{\psi}$ for a region $A$ of the chain is given by~\cite{stabilizer_entanglement}
\begin{equation}
\label{ent_stabilizer}
    S_A(\ket{\psi})=n_A-\log_2|G_A|.
\end{equation} 
Here $n_A$ is the number of qubits in $A$ and $|G_A|$ is the total number of stabilizers with support within $A$, i.e., of stabilizers that act trivially on qubits outside of $A$. Let $\mathcal{S} = \{s_1,\dots,s_{n}\}$ denote a generating set of the subgroup $G_A$, with $n=\log_2|G_A|$. Without loss of generality, we can assume that for every site $i\in A$, there are at most two stabilizers in the generating set that start or end at $i$ (i.e., for which the first or last nontrivial Pauli operator is at $i$). Each of the generators $s_i$ can be written as a product of $g$ and $Z$ operators contained in $A$ (for a contiguous region $A$). 

We now consider the set of operators $\Tilde{\mathcal{S}}=\{\tilde{s}_1, \dots, \tilde{s}_n\}$, where $Z$ operators are replaced by $g$ and vice versa. By the previous Lemma, these operators are stabilizers of the state of the dual circuit, and one can easily check that they are linearly independent with respect to the product of Pauli strings (since the original set is linearly independent). We define the region $\Tilde{A}=A$ in the $\alpha$ even case and $\Tilde{A} = \{i +1/2| i\in A\}$ in the $\alpha$ odd case. In order to obtain an upper bound for the entanglement entropy of $\Tilde{A}$, we find a lower bound for the number of stabilizers $\Tilde{s}_i$ fully contained inside $\Tilde{A}$. We observe that the transformation $Z_i\rightarrow g_i$ moves the first and last nontrivial operators by $\alpha/2$ positions to the left and to the right, respectively. Therefore, $\Tilde{s}_i$ is still fully contained inside $\Tilde{A}$ if $s_i$ does not have any nontrivial operator closer than $\alpha/2$ to one of the edges of $A$. Since there are at most two operators with initial or final nontrivial operator at each site inside $A$, at least $n - 2\alpha$ of the string operators in $\Tilde{\mathcal{S}}$ will be contained in $\Tilde{A}$. Therefore $\log_2|G_{\tilde{A}}| \geq \log_2|G_A| - 2\alpha$. Comparing to equation \eqref{ent_stabilizer} gives
\begin{equation}
    S_{\tilde{A}}\left(\ket{\Tilde{\psi}}\right) \leq S_A(\ket{\psi})+2\alpha,
\end{equation}
proving that the state of the dual circuit $\ket{\Tilde{\psi}}$ indeed obeys area law entanglement scaling for any area law entangled state $\ket{\psi}$ of the original circuit. This completes our proof that the critical point must be self-dual, $p_{s}^{{\rm crit}}=1/2$.

\section{Convergence of MPS simulations with bond dimension}
\label{appendix_B}
\setcounter{equation}{0}
\renewcommand{\theequation}{B\arabic{equation}}

For circuits with  Haar random unitary gates, we rely on the time-evolving block decimation (TEBD) algorithm to study the time evolution. In this method, the state of the system is represented as an MPS with maximal bond dimension $\chi_{\text{max}}$. Such an MPS captures the exact state in area law phases for large enough $\chi_{\text{max}}$, while volume law phases require a bond dimension that diverges with increasing system size in the thermodynamic limit~\cite{DMRG_MPS,tenpy}. In this Appendix, we examine the convergence of our numerical results with $\chi_{\text{max}}$ for the XZX Haar random circuit model in the different regions of the phase space. 

\begin{figure}[t!]

 \centering

  \includegraphics{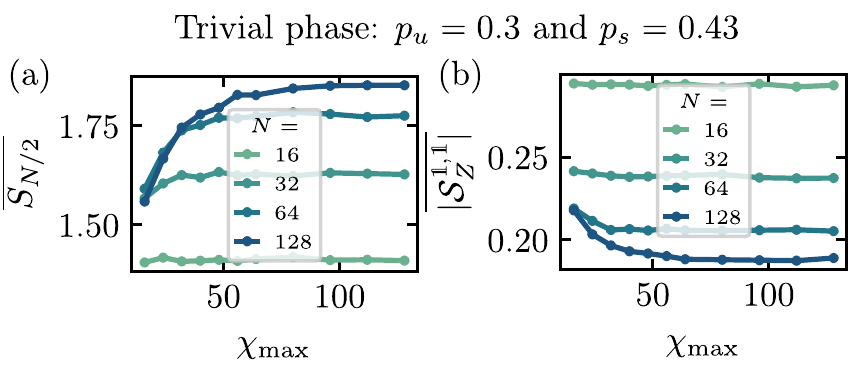}

  \caption{Convergence of TEBD results in the trivial area law phase of XZX Haar random circuit. (a) Half-chain entanglement entropy and (b) string order parameter $\overline{|\mathcal{S}_{Z}^{\mathds{1},\mathds{1}}|}$ as a function of the maximum bond dimension $\chi_{\text{max}}$ for different system sizes $N$ at $p_u=0.3$ and $p_s=0.43$, in the trivial phase. Points obtained by averaging over $10^3$ trajectories.}

  \label{fig:convergence_trivial}

\end{figure}

We first look at the behavior in the trivial area law phase. In Fig.~\ref{fig:convergence_trivial}a and b we show the convergence of the entanglement entropy and the string order parameter as a function of maximum bond dimension for the parameters $p_u=0.3$ and $p_s=0.43$. We observe that the entanglement entropy converges to a finite value as the system size is increased, confirming that the stationary state shows area law entanglement. Therefore, only a finite bond dimension is required to properly represent the state of the system at any time step.

\begin{figure}[t!]
    \centering
    \includegraphics{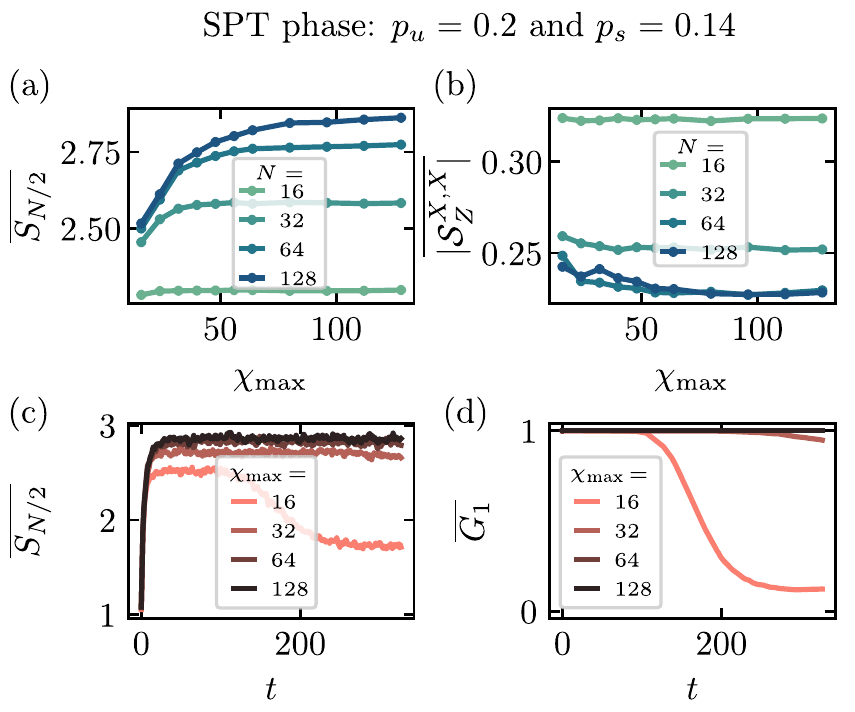}
    \caption{Convergence of TEBD results in the SPT phase of XZX Haar random circuit. Upper row: (a) Half-chain entanglement entropy and (b) string order parameter $\overline{|\mathcal{S}_{Z}^{X,X}|}$ as a function of the maximum bond dimension $\chi_{\text{max}}$ for different system sizes $N$ at $p_u=0.2$ and $p_s=0.14$ in the SPT phase. Bottom row: Time evolution of the (c) symmetry operator $G_1$ and (d) half-chain entanglement entropy $S_{N/2}$, for different bond dimensions and fixed system size $N=128$ in the SPT phase. The evolution is time-averaged over $10^3$ realizations of the circuit. Truncation performed in the TEBD algorithm can break the symmetry at low bond dimensions, leading to a reduced $S_{N/2}$.}
  \label{fig:bond_dim_spt}
\end{figure}
\begin{figure*}[t!]

 \centering

  \includegraphics{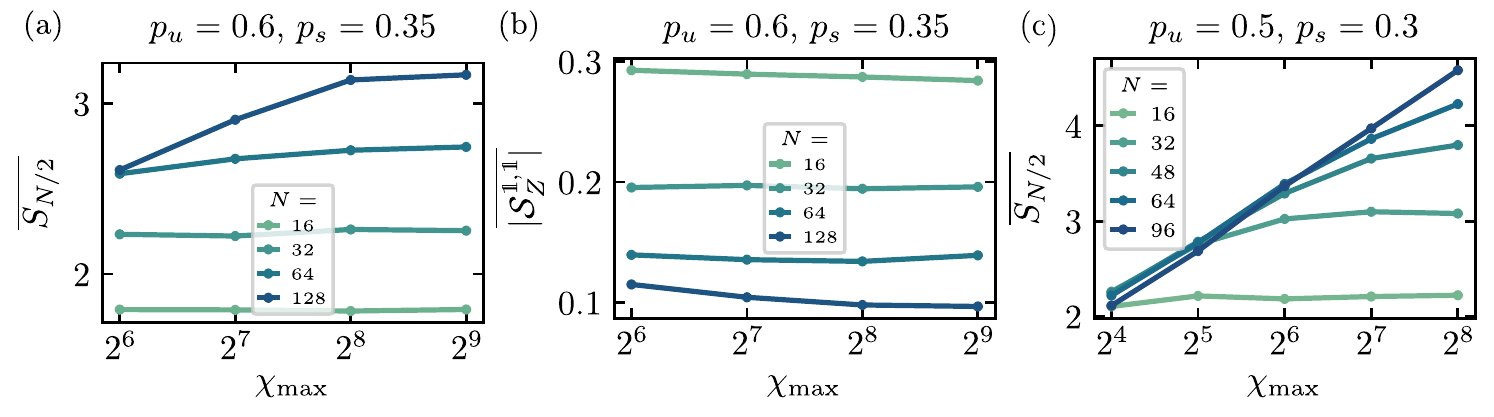}

  \caption{Dependence of the TEBD results on bond dimension $\chi_{\text{max}}$ close to the critical point and in the volume law phase of the XZX Haar random circuit. (a) Half-chain entanglement entropy and (b) string order parameter $\overline{|\mathcal{S}_{Z}^{\mathds{1},\mathds{1}}|}$ as a function of the maximum bond dimension $\chi_{\text{max}}$ for different system sizes $N$ at $p_u=0.6$ and $p_s=0.35$, in the vicinity of the numerically extracted critical point between the trivial phase and the volume law phase. Results consistent with a logarithmic scaling, $\overline{S_{N/2}}\sim \log N$. (c) Half-chain entanglement entropy as a function of bond dimension for a point inside the volume law phase ($p_u=0.5$ and $p_s=0.3$), where the required bond dimension increases exponentially with system size. Points obtained by averaging over $10^2$ trajectories.}

  \label{fig:convergence_critical_volume}

\end{figure*}
Ensuring convergence with bond dimension is trickier in the SPT phase, since truncating the MPS to bond dimension to $\chi_{\text{max}}$ can slightly break the $\mathbb{Z}_2\cross\mathbb{Z}_2$ symmetry of the model. This may happen when the Schmidt decomposition of the state has degenerate singular values, an effect already well-known from the MPS representation of SPT phases in equilibrium~\cite{ent_spectrum_Pollmann}. In particular, since the symmetry implies a 4-fold degenerate entanglement spectrum (see Appendix~\ref{appendix_D}),  truncating the MPS to a sufficiently small odd $\chi_{\text{max}}$ and thereby breaking this degeneracy always breaks the $\mathbb{Z}_2\cross\mathbb{Z}_2$ symmetry. Such a truncation error affects the averaged long-time values of certain quantities. More generally, in the circuit model, the entanglement spectrum can show even higher degeneracies at certain time steps, always in powers of 2. For this reason, it is convenient to always choose $\chi_{\text{max}}$  as a power of 2, leading to better conservation of the symmetry during the truncation step in the TEBD evolution, even for relatively small bond dimensions. We show the convergence of $S_{N/2}$ and the string order parameter with $\chi_{\text{max}}$ for various system sizes $N$ in Figs.~\ref{fig:bond_dim_spt}a and b~\footnote{In the cases where symmetry is broken, we only average over time steps where the steady state has already been reached, but the symmetry is not yet broken.}, at the point with $p_u=0.3$ and $p_s=0.14$, belonging to the  SPT phase. We obtain results similar to those of the trivial phase, where entanglement entropy and string order converge to a finite value for a fixed finite bond dimension, regardless of the system size. Figures~\ref{fig:bond_dim_spt}c and d show the time evolution of the circuit-averaged half-chain entanglement entropy $S_{N/2}$ and of the absolute value of the symmetry operator $G_1$, respectively, for various maximal bond dimensions with a fixed number of qubits $N=128$. The smallest bond dimension, $\chi_{\text{max}}=16$,  is not enough to preserve the symmetry during the time evolution, and we observe that the entanglement entropy is reduced upon breaking the symmetry.

Finally, we discuss the convergence at points close to criticality and in the volume law phase. Figures~\ref{fig:convergence_critical_volume}a and b show the convergence of the averaged steady-state half-chain entanglement entropy $S_{N/2}$ and the string order parameter $\overline{|\mathcal{S}_Z^{\mathds{1},\mathds{1}}|}$ close to the phase transition between trivial area law and volume law phase ($p_u=0.6$, $p_s=0.35$). In Fig.~\ref{fig:convergence_critical_volume}a we observe a constant increase of the entanglement entropy as the system size is doubled, indicating a logarithmic scaling of entanglement entropy. This implies that the bond dimension required for convergence scales as $\chi_{\text{max}}\propto N$. For sufficiently small system sizes $N\lesssim 128$, we are able to get converged results for the TEBD algorithm by taking a large enough bond dimension, up to $\chi_{\text{max}}=512$. This behavior is quite different from that obtained in the volume law phase, where bond dimension scales exponentially with system size. Therefore, no convergence is observed for the reachable bond dimensions, even for considerably smaller system sizes, see Fig.\ref{fig:convergence_critical_volume}c.

We note that in contrast to Clifford random circuits, in the presence of Haar random unitary gates the position of the phase boundary between area law and volume law entanglement scaling can vary with the index of the R\'enyi entropy. In particular,  it has been shown \cite{A10} that in a certain class of random quantum circuits, the measurement-induced phase transition between area law and volume law entanglement is located at different critical probabilities for Rényi entropies $n=0$ and $n\geq 1$, with $p_c^{n=0} \geq p_c^{n\geq 1}$. In this case, for measurement probabilities $p_c^{n\geq 1} \leq p \leq p_c^{n=0}$, one needs a bond dimension extensive in system size to exactly describe the state with an MPS, even though the von Neumann entanglement entropy follows an area law scaling. For such states, it can not be determined whether the MPS representation is able to efficiently approximate the exact state of the system, based on the entanglement scaling alone~\cite{Schuch_2008}. For the family of circuit models studied here, we only relied on the von Neumann entropy and string order parameters to detect the phases, and we have checked the convergence of both these quantities with bond dimension within the area law phases.  Both methods indicated the same critical lines. Further study would be required to check the faithfulness of the MPS representation.

\section{Finite-size corrections to string order parameters}
\label{appendix_C}
\setcounter{equation}{0}
\renewcommand{\theequation}{C\arabic{equation}}

To determine the critical lines using string order parameters, we examine their finite-size behavior for various system sizes, comparing it to a power law decay expected to hold at the critical point. This procedure allows us to extrapolate our results to the thermodynamic limit, and to determine the phase boundaries with higher numerical precision. In the thermodynamic limit, the area law phases are distinguished by the vanishing of exactly one of the string order parameters, while both of them become zero in the volume law phase. In more detail, we use the following numerical procedure. For a fixed $p_u$, we calculate the relevant string order parameter for different measurement rates $p_s$ close to the transition, for several system sizes $N$. Then, for each $p_s$, we fit the results with a power law function $f(N)=c N^{-\eta}+b$. We expect that at the critical point the string order parameter approaches zero as $ N^{-\eta}$, thus the finite-size results can be well fitted  with $\eta>0$ and $b=0$. We note that away from the critical point, this fitting of finite-size results breaks down.

\begin{figure}[t!]
    \centering
    \includegraphics{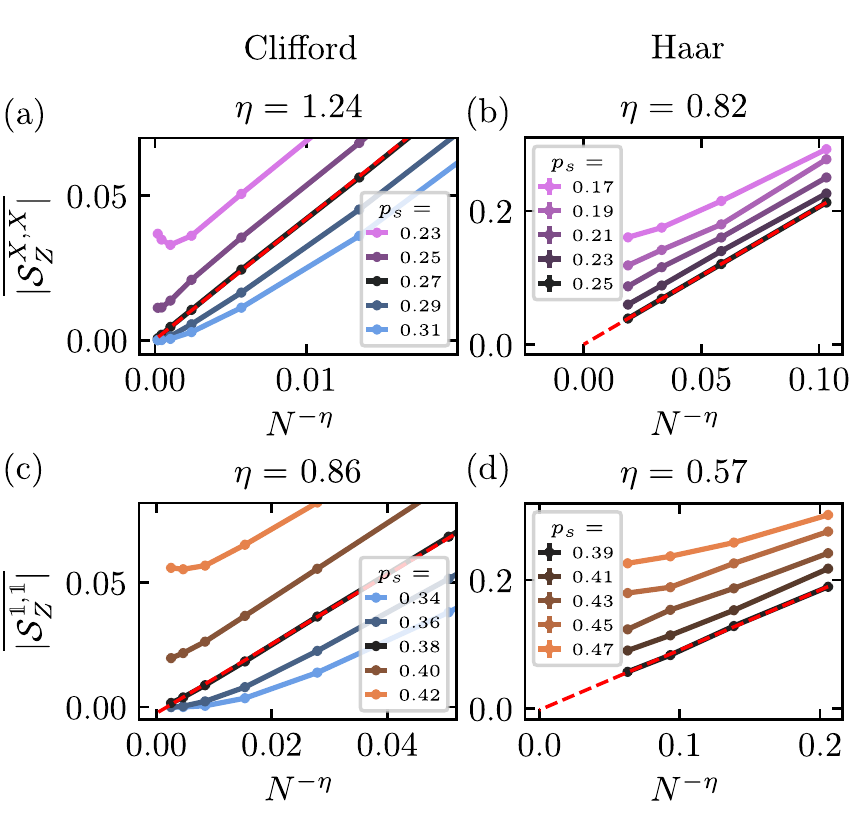}
    \caption{Finite-size corrections to string order parameters for the XZX circuit model at $p_u=0.2$. In (a) and (b) we plot $\overline{|\mathcal{S}_Z^{X,X}|}$ for various measurement rates $p_s$ across the SPT to volume law transition for Clifford and Haar random unitary evolution, respectively, as a function of $N^{-\eta}$. Similarly, in (c) and (d) we plot $\overline{|\mathcal{S}_Z^{\mathds{1},\mathds{1}}|}$ across the volume law to trivial area law transition. The exponent $\eta$ is obtained by linear regression at the $p_s$ value that gives the best fit of the form $N^{-\eta}$. The dashed red lines indicate the linear regression extrapolated to $N\rightarrow\infty$.}
  \label{fig:Cliff_strings_difNs}
\end{figure}

We illustrate this procedure for the XZX circuit model in Fig.~\ref{fig:Cliff_strings_difNs}, where we show the finite-size corrections to string order parameters at different single-qubit measurement rates $p_s$, for Clifford (left) and Haar (right) unitary gate rate $p_u=0.2$. We show the behavior of the string order parameter $\overline{|\mathcal{S}_Z^{X,X}|}$ for the transition between the SPT area law phase and the volume law phase (top), and the scaling of $\overline{|\mathcal{S}_Z^{\mathds{1},\mathds{1}}|}$ while crossing the boundary from the volume law to the trivial area law case (bottom).  In the Clifford case, for $\overline{|\mathcal{S}_Z^{X,X}|}$, the best fit of the form $N^{-\eta}$ is obtained for $p_s=0.27$, with $\eta=1.24$, see Fig.~\ref{fig:Cliff_strings_difNs}a. This result is consistent with the critical $p_s$ obtained based on entanglement entropies, separating the SPT area law and the volume law phases, see Ref.~\cite{Lavasani2021}. For slightly higher values of $p_s$, the string order parameter converges to zero even more rapidly, while for $p_s$ below the critical value, it remains finite in the thermodynamic limit. Similarly, Fig.~\ref{fig:Cliff_strings_difNs}c shows the Clifford results for $\overline{|\mathcal{S}_Z^{\mathds{1},\mathds{1}}|}$, predicting a transition at $p_s=0.38$ between the volume law and trivial area law phases, again consistent with the critical value found in Ref.~\cite{Lavasani2021}. Figures~\ref{fig:Cliff_strings_difNs}b and d show a similar extrapolation for Haar unitaries. In this case, the extrapolation has a larger error, due to the more limited system sizes accessible in MPS simulations. We observe a similar finite-size behavior for the XZZX Clifford model.

We note that in the area law phases the string order parameters converge nonmonotonously to a nonzero value for certain parameter sets; see, for example, the line for $p_s=0.23$ in Fig.~\ref{fig:Cliff_strings_difNs}a. This behavior is caused by the interplay of two different convergence rates of the string order parameter. Increasing the string length and the distance from the boundaries of the chain results in a convergence towards the order parameter value in the thermodynamic limit from up and from below, respectively, with the two processes characterized by different convergence rates. Monotonous convergence can be restored by fixing either of these length scales, but we found that a more involved numerical procedure accounting for this effect yields the same phase boundary within numerical precision.

\section{Entanglement spectrum of XZX Haar random circuit}
\label{appendix_D}
\setcounter{equation}{0}
\renewcommand{\theequation}{D\arabic{equation}}

The MPS representation of a state gives direct access to the Schmidt coefficients $\lambda_\alpha$ for any partition of the qubit chain into subsystems $A$ and $B$. These Schmidt values $\lambda_\alpha$ are  defined through
\begin{equation}
    \ket{\psi}_{AB}=\sum_\alpha \lambda_\alpha \ket{\psi_\alpha}_A\otimes\ket{\psi_\alpha}_B,
\end{equation}
where $\{\ket{\psi_\alpha}_{A}\}$ and $\{\ket{\psi_\alpha}_{B}\}$ are orthonormal bases for subsystems $A$ and $B$, respectively. Therefore, we have access to the entanglement spectrum of the system, defined as $-\ln\lambda_\alpha$, at different points of phase space. In this section, we focus on the entanglement spectrum of the half chain. We find equivalent results for any other nontrivial partition of the system. 

\begin{figure}[t!]
    \centering
    \includegraphics{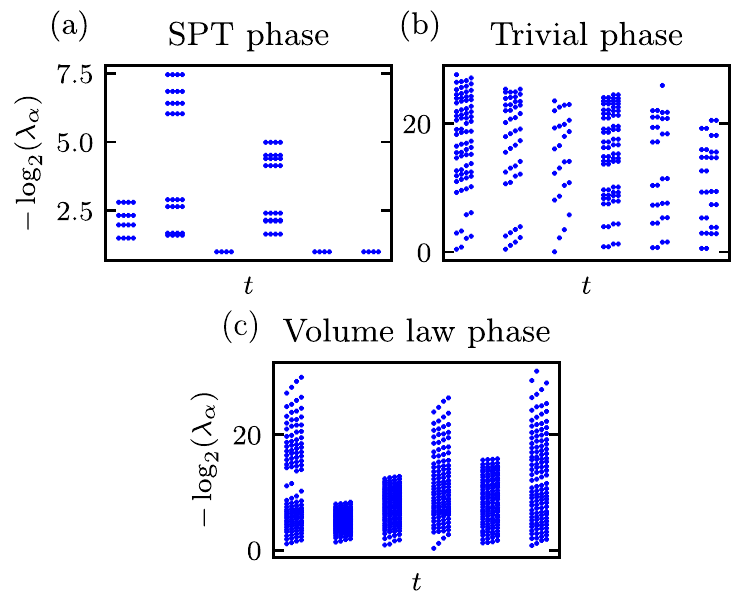}
    \caption{Entanglement spectrum of the XZX Haar random circuit model. The instantaneous entanglement spectrum is plotted for selected time steps in the steady state within the (a) SPT phase ($p_s=0.1$, $p_u=0.1$), (b) trivial phase ($p_s=0.45$, $p_u=0.1$) and (c) volume law phase ($p_s=0.6$, $p_u=0.25$). The entanglement spectrum shows a fourfold degeneracy in the SPT area law phase. The Schmidt values at the bottom, yielding the dominant contribution to the entanglement entropy, remain well separated in both area law phases, while forming a dense cluster in the volume law phase. We used  $N=256$ and $\chi_{\text{max}}= 128$.}
    \label{fig:entanglement_spectrum}
\end{figure}

In a nonequilibrium system, the entanglement spectrum changes with time. Therefore, we illustrate its structure in the steady state by showing the instantaneous entanglement spectrum at a few selected time steps. The results for the different phases of the model are displayed in Fig.~\ref{fig:entanglement_spectrum}.

In the SPT area law phase, displayed in Fig.~\ref{fig:entanglement_spectrum}a, we find that each Schmidt coefficient is fourfold degenerate. A similar fourfold degeneracy of the entanglement spectrum in the SPT phase with $\mathbb{Z}_2\cross\mathbb{Z}_2$ symmetry has been observed in equilibrium quantum matter~\cite{ent_spectrum_Pollmann}, a property that translates to this nonequilibrium setting. In the trivial area law phase, the fourfold degeneracy is lifted, but the Schmidt values at the bottom, most relevant for the entanglement entropy, remain well separated from the rest, see Fig.~\ref{fig:entanglement_spectrum}b.  Finally, in the volume law phase of Fig.~\ref{fig:entanglement_spectrum}c,  we find a large number of Schmidt values at the bottom of the spectrum, a sign of a highly entangled state. In this case, choosing a larger $\chi_{\text{max}}$ would add relevant Schmidt values, significantly changing the value of the entanglement entropy, as was discussed in Appendix~\ref{appendix_B}.

\section{Finite-size scaling with string order parameters}
\label{appendix_E}
\setcounter{equation}{0}
\renewcommand{\theequation}{E\arabic{equation}}

In Appendix~\ref{appendix_C} we have shown that string order parameters decay as a power law at the critical point with a critical exponent $\eta$. To further characterize the phase transitions and string order parameters discussed in this work, we examine a finite-size scaling ansatz for the string order parameters, and extract the critical exponent associated with the diverging correlation length $\nu$. In the measurement-only regime ($p_u=0$) of the XZX model,  $\nu=4/3$ has been established through a mapping to a percolation model~\cite{Lavasani2021}. For a nonzero rate of unitary evolution, the numerical results of Ref.~\cite{Lavasani2021} are consistent with $\nu=4/3$ for both phase boundaries, up to numerical precision. Here the exponent $\nu$ is obtained from a finite-size scaling ansatz for entanglement entropies.

We find that our numerical results are consistent with the following finite-size scaling ansatz for the string order parameters near the critical point~\cite{finite_scaling_sop},
\begin{equation}
\label{scaling_ansatz}
    \mathcal{S}_\Sigma^{O^L, O^R}(p, N) = N^{-\eta} F\left((p-p_c)N^{1/\nu_s}\right),
\end{equation}
with critical exponents $\nu_s$ and $\eta$. Note that, in principle, the exponent $\nu_s$  could differ from the  exponent $\nu$ obtained through entanglement entropies (see, for example, the different critical exponents for the entanglement entropy and the spin glass order parameter in Ref.~\cite{Sang_2021}). 

\begin{figure}[t!]
    \centering
    \includegraphics{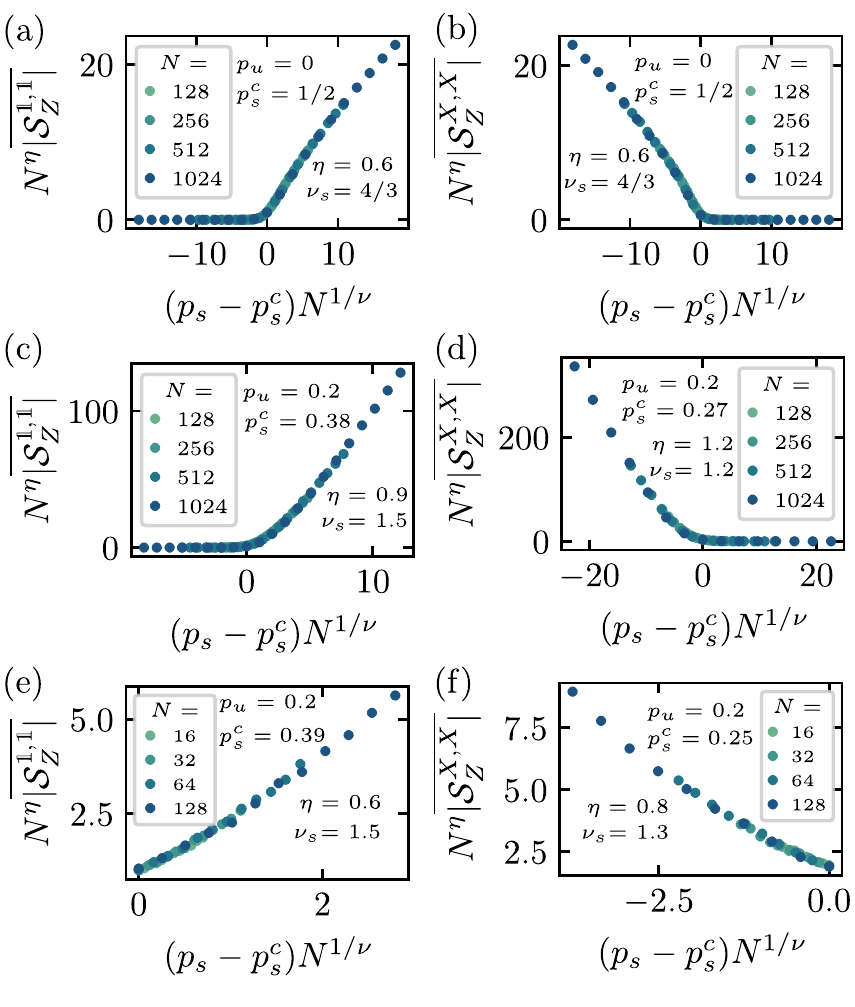}
    \caption{Finite-size scaling of the string order parameters in the XZX circuit cluster model. (a) and (b) show the collapse in the measurement-only limit in the vicinity of the critical point between the SPT and trivial area law phases, $p_c=0.5$. The critical exponents are $\eta\approx0.6$ and $\nu_s=4/3$. In the remaining panels (c)-(f), the rate of unitaries is set to $p_u=0.2$, and we display  the collapse for the transition from SPT area law to volume law (left), and from volume law to trivial area law (right) phase. (c) and (d) show the collapse  for the evolution with Clifford unitaries, while (e) and (f) display it for the circuit with random Haar evolution, obtained with MPS simulations. In the latter case, data for the string order are only available in the area law phase, where our simulations have converged with the available bond dimensions.}
  \label{fig:finite_scaling}
\end{figure}

We first analyze the scaling collapse of our data for the measurement-only evolution, $p_u=0$, where both the exact position of the critical point and the critical exponent $\nu=4/3$ are known  from a mapping to percolation. The scaling ansatz~\eqref{scaling_ansatz}  yields an excellent collapse with $\nu_s=4/3$ and $\eta\approx 0.6$ for both string order parameters, as shown in Figs.~\ref{fig:finite_scaling}a and b.  

When we turn the unitary evolution on, obtaining the parameters for the scaling collapse becomes more challenging. Here we need to find the best fit for three unknown parameters: the critical probability $p_s^c$ and the critical exponents $\eta$ and $\nu_s$. To perform the fitting, we set $p_s^c$ and $\eta$ to the values obtained in Appendix~\ref{appendix_C}, so that we only need to fit $\nu_s$ to get the best collapse. In Fig.~\ref{fig:finite_scaling}c and d we show the scaling collapse for the two transitions in the line $p_u=0.2$ for the evolution with Clifford unitaries. In Figs.~\ref{fig:finite_scaling}e and f we show it for random Haar unitaries, but limited to values in the area law phase, where our MPS simulations have converged with respect to bond dimension.

We note that, due the large number of fitting parameters, the obtained exponents suffer from considerable numerical uncertainty. Nevertheless, we can extract several conclusions from the scaling collapses. In both the Clifford and Haar cases, the values of $\nu_s$ that provide the best fitting are always in the interval $[1.1, 1.5]$, which could be consistent with the percolation value $4/3$ within numerical errors. We would need further large-scale simulations and a  more precise analysis of data to determine $p_s^c$ and $\eta$ with higher precision, and to carefully check  the consistency with the percolation value.  We leave this challenging task for future work. The good collapse that we obtain for the Haar random case is a clear indication of criticality, which reinforces our estimation of the phase transition points and the existence of a phase transition.

\begin{figure}[t!]
    \centering
    \includegraphics{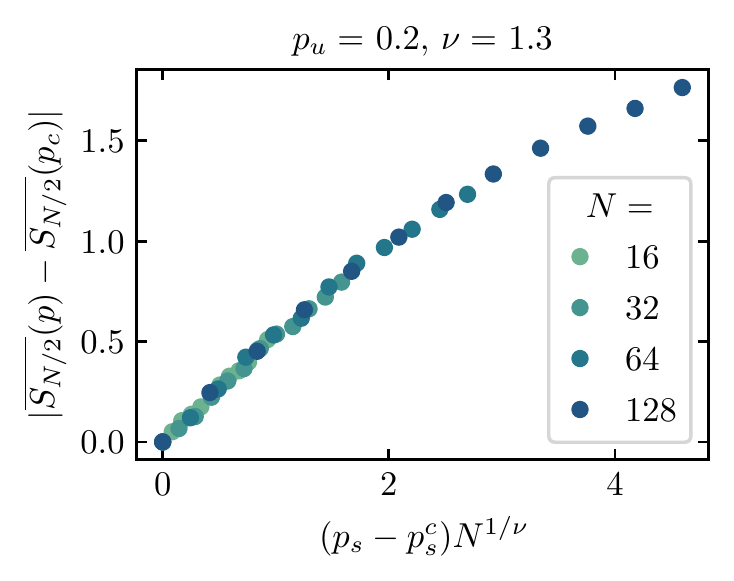}
    \caption{Finite-size scaling collapse of the disorder-averaged half-chain entanglement entropy, subtracting the value at the critical point. The exponent $\nu\approx1.3$ gives the best possible collapse for the available data.}
    \label{fig:finite_scaling_entropy}
\end{figure}

In the transition between trivial area law and volume law entanglement, one can perform a  similar finite-size scaling analysis for the half-chain entanglement entropy as discussed in Refs.~\cite{EPT3, Li_2019}. This scaling collapse relies only on the values of $p_s^c$ and $\nu$, allowing to perform the fitting with slightly less numerical uncertainty. The half-chain entanglement entropy is a quantity that can be obtained straightforwardly from MPS simulations. Results are shown in Fig.~\ref{fig:finite_scaling_entropy}, where we find a scaling exponent consistent with the percolation value $\nu=4/3$. Such a result extends to the rest of the trivial-volume law critical line. With these results, we show the suitability of MPS simulations to determine critical exponents in the context of measurement-induced phase transitions.

\newpage
\end{appendix}

\bibliographystyle{apsrev4-2}
\bibliography{references}

\end{document}